# ΘΕΩΡΙΑ ΚΑΙ ΠΕΙΡΑΜΑΤΑ ΣΤΟΝ ΚΒΑΝΤΙΚΟ ΝΟΥ

Ανδρέας Μέρσιν [1] και Δημήτρης Β. Νανόπουλος[1,2,3]

Στο άρθρο αυτό δίνουμε μια περίληψη του μοντέλου εγκεφαλικής λειτουργίας που αναπτύξαμε αναγνωρίζοντας τους κβαντομηχανικούς μηχανισμούς οι οποίοι πιθανότατα παίζουν πρωταρχικό ρόλο στην λειτουργία του νευρώνα (Νανόπουλος κ.α. 1999) Αναφέρουμε επίσης τα αποτελέσματα των προκαταρκτικών *in vivo* πειραμάτων μας με την γενετικά τροποποιημένη μύγα *Δροσόφυλλα* και περιγράφουμε τα επόμενα *in vitro* πειράματα που θα πραγματοποιήσουμε. Με βάση την θεωρητική μας μελέτη των μικροσωληνιδίων, προτείνουμε μηχανισμούς για την μοριακή βάση του "εγγράμματος" της μνήμης όπου θεωρούμε την πρωτεΐνη τουμπουλίνη με τις εναλλασσόμενες καταστάσεις διπολικής ροπής της, ως την βασική μονάδα ένδο-νευρωνικού υπολογισμού και τον κυτταροσκελετό ως την υποδομή της μνήμης.

[1] Τμήμα Φυσικής, Πανεπιστήμιο Τέξας A&M, Η.Π.Α.
[2] Ακαδημαϊκός, Πρόεδρος Τμήματος Φυσικών Επιστημών, Ακαδημία Αθηνών
[3] Head, Astroparticle Physics Group, Houston Advanced Research Center (HARC)

E-mail: mershin@physics.tamu.edu

## 1. Εισαγωγή

Εδώ και τρεις δεκαετίες τουλάχιστον, θεωρητικοί φυσικοί έχουν προτείνει ότι η λειτουργία μακρομοριακών βιολογικών συστημάτων όπως π.χ. των πρωτεϊνών (Froelich 1983) σχετίζεται άμεσα με τις αρχές της κβαντομηχανικής, μιας επιστήμης που συνήθως περιγράφει τα φαινόμενα του ατομικού και υποατομικού κόσμου. Η αίσθηση της πλειοψηφίας της βιολογικής κοινότητας ήταν, ότι ακόμη και αν κβαντικά φαινόμενα όντως συναντούνται σε βιολογικά συστήματα, δεν παίζουν κάποιο σημαντικό ρόλο. Αυτό έχει ως αποτέλεσμα, μέχρι πριν δέκα χρόνια περίπου, το ενδιαφέρον στην κβαντική προσέγγιση βιολογικών συστημάτων να έχει παραμείνει ως επί το πλείστον μαθηματικό. Το 1989, ο παγκοσμίου φήμης Άγγλος μαθηματικός και θεωρητικός φυσικός Roger Penrose δημοσίευσε ένα βιβλίο με τίτλο "Το καινούργιο μυαλό του αυτοκράτορα" (Penrose 1989) στο οποίο έθεσε το ερώτημα εάν η κλασσική (μη-κβαντική) φυσική, και κατ' επέκταση όλες οι βιολογικές επιστήμες μπορούν να ελπίζουν ότι κάποια μέρα θα δώσουν απάντηση σε ερωτήματα όπως π.χ. πώς δημιουργούνται οι ιδέες, η έμπνευση ή η κατανόηση της αλήθειας μιας μαθηματικής απόδειξης. Ο Penrose επιχείρησε να δείξει ότι καμία από τις σημερινές τεχνολογικές προσπάθειες εξομοίωσης του εγκεφάλου ως εξαιρετικά εξελιγμένου μεν, κοινού (κλασικού) δε υπολογιστή δεν θα μπορέσει να δημιουργήσει μηχανές με πραγματική συνείδηση. Συνεχίζοντας το επιχείρημά του, το οποίο βάσισε πάνω στη μαθηματική ιδέα της υπολογιστικής μηχανής Touring, κατέληξε στο συμπέρασμα ότι η συνείδηση είναι μία εκδήλωση της κβαντικής φύσης του εγκεφάλου, ως πηγή της οποίας πρότεινε τα μικροσωληνίδια των εγκεφαλικών νευρώνων. Τέλος, ανέπτυξε την ιδέα ότι ακόμα και οι βασικοί νόμοι της φυσικής πρέπει να αναθεωρηθούν και μόνο όταν υπάρξει μια θεωρία κβαντικής βαρύτητας θα μπορέσουν οι φυσικοί και βιολόγοι να καταλάβουν επαρκώς τη λειτουργία του νου ώστε να μπορέσουν να τον μιμηθούν. Όπως ήταν αναμενόμενο, και σωστά κατά την αντίληψή μας, η επιστημονική κοινότητα δέχτηκε σχετικά ψυχρά τις ιδέες αυτές με κύριο λόγο όχι μόνο την πλήρη απουσία πειραματικών δεδομένων αλλά και την φαινομενική έλλειψη οποιασδήποτε πιθανότητας πειραματικής εξέτασης και επαλήθευσής τους.

Εντούτοις, λόγω της σπουδαιότητας και ενδιαφέροντος του ερωτήματος, καθώς επίσης και της ειδικότητας της ομάδας μας στην κβαντομηχανική, θεωρήσαμε ότι οι ιδέες αυτές είναι χρήσιμο να εξεταστούν όσο το δυνατόν πιο εξαντλητικά ως προς την ορθότητά τους, (Νανόπουλος 1994, Mavromatos κ.α. 2002) και σχετικά πειράματα πρέπει να προταθούν και να εκτελεσθούν. Κατά την διάρκεια της έρευνάς μας, οδηγηθήκαμε σε μια διαφορετική αντίληψη απ' αυτή του Penrose, και προτείναμε ότι η κβαντική φύση του μικρόκοσμου, υπό συγκεκριμένες συνθήκες, μπορεί να επεκταθεί στον μακρόκοσμο όλων των κυττάρων, καθώς επίσης και ολόκληρων συστημάτων όπως ο εγκέφαλος και μπορεί να ανιχνευτεί κάνοντας προσεχτικά σχεδιασμένα φυσικά και βιολογικά πειράματα. Επίσης, ανακαλύψαμε ότι η προσπάθεια εξήγησης της λειτουργίας των πρωτεϊνών αλλά και συστημάτων όπως αυτό της μνήμης διευκολύνεται με την χρήση της κβαντικής φυσικής και πολλά προς το παρόν ανεξήγητα φαινόμενα όπως αυτό του binding εξηγούνται.

Εδώ παρουσιάζουμε μία περίληψη της θεωρητικής μας προόδου, τα αποτελέσματα των προκαταρκτικών βιολογικών πειραμάτων που προτείναμε και εκτελέσαμε όπως επίσης και μια περιγραφή του επόμενου κύκλου πειραμάτων που θα πραγματοποιήσουμε.

## 2. Προβλήματα σημερινής αντίληψης περί μνήμης και εγκεφαλικής λειτουργίας

Η μάθηση και η μνήμη εκδηλώνονται ως διαφοροποιήσεις της αντίδρασης ενός συστήματος σε περιβαλλοντολογικά και εσωτερικά ερεθίσματα και αντικατοπτρίζονται στην λειτουργία του εγκεφάλου. Παρόλο ότι είναι γενικά αποδεκτό ότι αλλαγές στην βιοχημική κατάσταση των νευρώνων, και ιδιαίτερα των συνάψεών τους, μεσολαβούν στην εγγραφή μνήμης και στην επακόλουθη αλλαγή της λειτουργίας του εγκεφάλου, και υπάρχει πολύ ικανοποιητική κατανόηση διαφόρων τρόπων επικοινωνίας μεταξύ νευρώνων, αυτή την στιγμή δεν υπάρχει πλήρης κατανόηση του τρόπου με τον οποίο αντιδράσεις σε μοριακό επίπεδο επηρεάζουν αυτά τα συμβάντα. Μία από τις μεγαλύτερες προκλήσεις της σημερινής νευροβιολογίας είναι η εξήγηση της "κατακερματισμένης αλλά ολοκληρωμένης" φύσης της μνήμης, με άλλα λόγια το γεγονός ότι όπως έχουν δείξει πειραματικές μελέτες, κατά την διάρκεια μίας ανάμνησης, νευρώνες από διάφορα, μακροσκοπικά απομακρυσμένα μέρη του εγκεφάλου ενεργοποιούνται *ταυτόχρονα* και δημιουργούν την αίσθηση "ολοκληρωμένης" μνήμης. Επίσης, η τεράστια χωρητικότητα και ταχύτητα της ανθρώπινης μνήμης είναι δύσκολο να εξηγηθεί με μόνη βάση την ανταλλαγή νευροδιαβιβαστών μεταξύ των $10^{10}$ νευρώνων μεσώ των $10^{14}$ συνάψεων. Ακόμη κι αν κάποιος λάβει υπ' όψιν ηλεκτροτονικές επαφές στους κόμβους Ranvier, η σύγχρονη ενεργοποίηση μεγάλων αριθμών νευρώνων και <u>η</u> χωρητικότητα της μνήμης δεν εξηγείται.

## 3. Νευρώνες, τουμπουλίνη και μικροσωληνίδια

Ο κυτταροσκελετός των ευκαριωτικών κυττάρων αποτελείται κυρίως από νηματοειδή μικροσυμπλέγματα (microfilaments), ακτίνη και μικροσωληνίδια που αποτελούνται από τουμπουλίνη. Πρωτεΐνες ανάλογες με την ανθρώπινη τουμπουλίνη βρίσκονται ιδιαίτερα εμπλουτισμένες στους εγκεφαλικούς ιστούς όλων σχεδόν των ζώων, οι οποίοι περιέχουν νευρώνες με εξαιρετικά επιμήκη και διατεταγμένα μικροσωληνίδια που δίνουν το χαρακτηριστικό μήκος στους νευρωνικούς άξονες. Οι πιο κοινές μορφές τουμπουλίνης λέγονται α- και β - τουμπουλίνη και είναι σφαιροειδή πανομοιότυπα μονομερή με μοριακό βάρος περίπου 55,000 Dalton και διαστάσεις 46x65x40 Angstrom. Σε φυσιολογικές *in vivo* συνθήκες, η τουμπουλίνη βρίσκεται ως διμερές αποτελούμενο από α- και β- τουμπουλίνη και τείνει να δημιουργεί αυτογενή πολυμερή νημάτια τα οποία οργανώνονται (συνήθως ανά 13) με σχήμα κούφιου σωλήνα εσωτερικής και εξωτερικής διαμέτρου 14nm και 25nm αντίστοιχα τον οποίο ονομάζουμε μικροσωληνίδιο (βλ. εικόνα 1). Τα μικροσωληνίδια βρίσκονται σε εξαιρετικά οργανωμένη και σταθερή κατάσταση στον νευρωνικό άξονα όπως φαίνεται από φωτογραφίες μεσώ οπτικού και ηλεκτρονικού μικροσκοπίου (βλ. εικόνα 2) όπου επίσης φαίνονται μόρια πρωτεΐνης "tau" ή "τ" ως συνδετικές γέφυρες (Microtubule Associated Proteins -MAPs). Οι διάφορες λειτουργίες των μικροσωληνιδίων αφορούν στην μίτωση[1], κίνηση, ανάπτυξη, αλλαγή και διατήρηση σχήματος των κυττάρων, και ενδοκυτταρική μεταφορά ουσιών. Στην συγκεκριμένη περίπτωση, τουλάχιστον των εγκεφαλικών νευρώνων, προτείνουμε επίσης ότι τα μικροσωληνίδια εκτελούν και κβαντική επεξεργασία δεδομένων.

## 4. Κβαντική φυσική και κβαντικοί υπολογιστές

Η κβαντική φυσική στηρίζεται σε ένα από τα σπουδαιότερα και ακριβέστερα μαθηματικά πλαίσια της σημερινής επιστήμης, την κβαντομηχανική. Πολλές φορές, η κβαντομηχανική περιγράφεται ως μυστηριώδης και περίεργη λόγω της μη-διαισθητικής της φύσης. Για παράδειγμα, στην γλώσσα της κβαντομηχανικής ένα σωμάτιο μπορεί να βρίσκεται σε κατάσταση "επαλληλίας" όπου, εάν δεν 'παρατηρηθεί' βρίσκεται σε δύο ή περισσότερα μέρη ταυτόχρονα. Ο λόγος, κατά την άποψή μας, για την αδυναμία του ανθρώπου να αισθάνεται άνετα με τις περίεργες έννοιες της κβαντομηχανικής αλλά και άλλων απόψεων της σύγχρονης φυσικής, (όπως π.χ. οι 26 διαστάσεις της θεωρίας των υπερχορδών), είναι βιολογικός-ψυχολογικός: ποτέ στο παρελθόν κατά την διάρκεια της εξέλιξής μας δεν ήταν απαραίτητο ή χρήσιμο να κατανοούμε έννοιες όπως η επαλληλία καταστάσεων ή να σκεφτούμε σε παραπάνω από τέσσερις διαστάσεις (3 χώρου και 1 χρόνου). Η μη-διαισθητικότητα της κβαντικής μηχανικής δεν την κάνει ωστόσο λιγότερο ακριβής και έτσι αναγκαζόμαστε να την χρησιμοποιούμε όταν η φύση το απαιτεί, δηλαδή στα ατομικά και υποατομικά φαινόμενα αλλά καθώς φαίνεται, και στα φαινόμενα του εγκεφάλου.

---

[1] Η μίτωση αναφέρεται συνήθως ως προς την λειτουργία των μικροσωληνιδίων σε μη-νευρικά κύτταρα, καθώς θεωρείται ότι οι νευρώνες δεν διαιρούνται. Πρόσφατες μελέτες όμως, έχουν δείξει ότι τουλάχιστον μερικοί τύποι εγκεφαλικών νευρώνων πολλαπλασιάζονται καθ' όλη την διάρκεια της ζωής.

Οι κβαντικοί υπολογιστές είναι μια σχετικά νέα ιδέα που προτάθηκε πρώτα από τον Νομπελίστα Αμερικάνο φυσικό Richard Feynman (Feynman 1981) την δεκαετία του 1980. Ο Feynman πρότεινε ότι μπορούμε να κατασκευάσουμε υπολογιστές που αντί να αποτελούνται από κλασσικά τρανζίστορ όπου τα "μπιτ" μπορούνε να βρίσκονται στις δυαδικές καταστάσεις "0" και "1", να αποτελούνται από κβαντικά "μπιτ" που λόγω της αρχής της επαλληλίας θα μπορούν να βρίσκονται στις κλασσικές καταστάσεις "0" και "1" αλλά επίσης και σε άπειρες άλλες καταστάσεις επαλληλίας μεταξύ "0" και "1" ταυτόχρονα. Τα πλεονεκτήματα ενός τέτοιου υπολογιστή είναι τεράστια. Έχει, για παράδειγμα, αποδειχτεί ότι όταν κατασκευαστούν τέτοιες μηχανές θα μπορούν να λύνουν συγκεκριμένα διαβόητα μαθηματικά προβλήματα όπως αυτό της παραγοντοποίησης μεγάλων ακεραίων αριθμών ή της αναζήτησης στοιχείων σε μία βάση δεδομένων σε ελάχιστο χρόνο. Η εξαιρετικά γρήγορη λειτουργία τέτοιων υπολογιστικών μηχανών θα βοηθήσει στην επίλυση προβλημάτων όπως αυτό της κατανόησης του γονιδιόματος και αυτό του protein folding. Αυτό όμως που κάνει τους κβαντικούς υπολογιστές του μέλλοντος ακόμα πιο ενδιαφέροντες δεν είναι μόνο η δυνατότητα αποθήκευσης γιγαντιαίων αριθμών δεδομένων σε ελάχιστο χώρο, ούτε η τεράστια ταχύτητα υπολογισμού. Οι κβαντικοί υπολογιστές θα έχουν τελείως νέο τρόπο λειτουργίας ο οποίος θυμίζει έντονα τον τρόπο με τον οποίο σκέφτεται ο άνθρωπος. Με άλλα λόγια, αυτό που συμβαίνει είναι ότι ο κβαντικός υπολογιστής "κατασταλάζει" σε μία απόφαση μετά από μια (μικρή) περίοδο κατά την οποία τα δεδομένα και οι υπολογισμοί είναι σε μία κατάσταση κβαντικής επαλληλίας, σε αναλογία με την γέννηση μιας ιδέας στον ανθρώπινο νου, πριν τη δημιουργία της οποίας .διάφορες αδιαμόρφωτες ιδέες, γνώσεις, σχετικά και μη δεδομένα φαίνονται να είναι σε κατάσταση επαλληλίας.

Το πρόβλημα που συναντάται στην κατασκευή πρακτικά χρήσιμων κβαντικών υπολογιστών σήμερα, είναι ότι η κατασκευή κατάλληλων κβαντικών "μπιτ" είναι δύσκολη, διότι κατά την κβαντομηχανική ο χρόνος για τον οποίο οποιοδήποτε αντικείμενο βρίσκεται σε κατάσταση επαλληλίας είναι αντιστρόφως ανάλογος με την μάζα του αντικειμένου και την θερμοκρασία του περιβάλλοντος στο οποίο βρίσκεται. Έτσι, οι προσπάθειες κατασκευής κβαντικών υπολογιστών έχουν επικεντρωθεί σε ατομικά συστήματα χαμηλής θερμοκρασίας.

### 5. Το μοντέλο μας για το κβαντικό μπιτ και το έγγραμμα

Αυτό που πρότεινε η ομάδα μας είναι ότι υπάρχουν ήδη στην φύση κβαντικά μπιτ και κβαντικοί υπολογιστές οι οποίοι λειτουργούν σε θερμοκρασία δωματίου. Με βάση την κβαντική ηλεκτροδυναμική και γενικότερα την κβαντική θεωρία πεδίου, αναλύσαμε την δομή των μικροσωληνιδίων και της πρωτεΐνης τουμπουλίνης και αναγνωρίσαμε ότι η ηλεκτρική διπολική ροπή της πρωτεΐνης αυτής, υπό συγκεκριμένες συνθήκες όπως αυτές στο εσωτερικό ενός ζωντανού κυττάρου, μπορεί να βρίσκεται σε επαλληλία δύο καταστάσεων δίνοντας μας έτσι ένα κβαντικό μπιτ. Επίσης η ανάλυση έδειξε ότι μικροσωληνίδια και ακόμη και ολόκληροι νευρώνες είναι δυνατόν να βρίσκονται σε κατάσταση επαλληλίας για χαρακτηριστικούς χρόνους συγκρίσιμους με αυτούς νευροβιολογικών συστημάτων. Έτσι εξηγείται η φύση της ολοκληρωμένης μνήμης, καθώς νευρώνες που διαθέτουν πληροφορίες που ανήκουν στην ίδια μνήμη, π.χ. το σχήμα του προσώπου και ή χροιά της φωνής ενός γνωστού, δεν χρειάζεται να ενεργοποιηθούν εξατομικευμένα με αργά-κινούμενους νευροδιαβιδαστές, παρά φτάνει να λάβει σήμα ένας και όλοι οι νευρώνες που βρίσκονται σε επαλληλία με αυτόν ενεργοποιούνται ταυτοχρόνως. Η κβαντική θεωρία εγκεφάλου δεν διίστανται με καμία από τις καλά εμπεριστατωμένες παρατηρήσεις και ανακαλύψεις της νευροβιολογίας του τελευταίου αιώνα, απλά δίνει στους νευρώνες την δυνατότητα να "καλούν υπεραστικά" και γρήγορα άλλους σχετικούς νευρώνες έτσι ώστε να μπορεί να αρχίσει η διαδικασία κλασσικής ταυτόχρονης λειτουργίας μερών του εγκεφάλου, όπως παρατηρείται στα σημερινά πειράματα fMRI (functional Magnetic Resonance Imaging), και PET (Positron Emission Tomography).

Το "έγγραμμα" της μνήμης στο μοντέλο αυτό συνίσταται από την διάταξη των πρωτεϊνών MAPs που γεφυρώνουν τα μικροσωληνίδια, φτιάχνοντας κόμβους οι οποίοι σε αναλογία με τα δάκτυλα του κιθαρίστα, αναγκάζουν τα μικροσωληνίδια να έχουν διαφορετικές χαρακτηριστικές συχνότητες αλληλεπίδρασης με άλλα μικροσωληνίδια ή και νευρώνες.

## 6. Πειράματα

Παρόλο που η θεωρητική ανάλυση έδειξε ότι η τουμπουλίνη και κατ' επέκταση τα μικροσωληνίδια μπορούν να παίξουν ρόλο κβαντικής μνήμης και επεξεργαστή, πειράματα στον τομέα αυτόν δεν υπήρξαν ποτέ.

Ο πρώτος κύκλος πειραμάτων μας, αποσκοπούσε στην επιβεβαίωση της θεωρητικής μας πρόβλεψης ότι τα μικροσωληνίδια είναι άμεσα συνδεδεμένα με την λειτουργία και αποθήκευση της μνήμης. Για τον λόγο αυτόν, συνεργαστήκαμε με τον καθ. Βιολογίας Ευθ. Σκουλάκη (Νανόπουλος κ.α. 1999) και χρησιμοποιώντας τις μεθόδους τις οποίες είχε αναπτύξει για την ποσοτική ανάλυση της αποθήκευσης μνήμης στην μύγα *Δροσόφυλλα,* βρήκαμε ότι μύγες γενετικά τροποποιημένες έτσι ώστε τα μικροσωληνίδια τους να μην είναι άρτια οργανωμένα, παρόλο που απ' όλες τις άλλες απόψεις δεν διαφέρουν από φυσιολογικές μύγες "wild type", πάσχουν από αμνησία (Μέρσιν 2002). Αυτά τα πειράματα έδειξαν ξεκάθαρα για πρώτη φορά ότι τα μικροσωληνίδια είναι όντως σημαντικά στοιχεία του μηχανισμού της μνήμης. Δείξαμε επίσης ότι η κβαντική θεωρία εγκεφάλου όχι μόνο επιδέχεται πειραματική επαλήθευση αλλά δύναται και να παράγει προβλέψεις άμεσου ενδιαφέροντος στους νευροβιολόγους.

Ο δεύτερος κύκλος πειραμάτων μας, αποσκοπεί στην χρήση μικροσωληνιδίων για την συνθετική κατασκευή πλακιδίων ολοκληρωμένων κυκλωμάτων (chips) που θα μιμούνται την λειτουργία του κυτταροσκελετού του νευρώνα εγγράφοντας δεδομένα στην διάταξη των MAPs. Χρησιμοποιώντας ακτίνες λέιζερ συγκεκριμένων μηκών κύματος μπορούμε να ενεργοποιήσουμε δίκτυα μικροσωληνιδίων και να παρατηρήσουμε εάν καταστάσεις επαλληλίας όντως διέπουν τέτοιες διατάξεις πρωτεϊνών.

**7. Επίλογος**

Από την μαθηματική μας ανάλυση αλλά και την πειραματική μας δραστηριότητα, συμπεραίνουμε ότι η κβαντική φυσική ενδέχεται να παίξει σημαντικό ρόλο στη νευροβιολογία του μέλλοντος βοηθώντας στην περαιτέρω κατανόηση των μοριακών αλληλεπιδράσεων που συμβαίνουν στο εσωτερικό του νευρώνα. Κατ' επέκταση, φαίνεται ότι σταδιακά θα βρούμε ότι τα φαινόμενα που τώρα θεωρούνται ότι ανήκουν αποκλειστικά στον ατομικό κόσμο, έχουν στην πραγματικότητα *και* μακροσκοπική υπόσταση όπως έδειξαν πρόσφατα πειράματα μακροσκοπικής επαλληλίας (Julsgaard κ.α. 2001).

Δεν είναι άλλωστε απίστευτο ότι εφόσον η κβαντική επεξεργασία δεδομένων προσφέρει πλεονεκτήματα, μόρια και κύτταρα εξειδικευμένα σε αυτήν θα έχουν εξελιχθεί κατάλληλα έτσι ώστε να εκμεταλλεύονται πλήρως την κβαντική τους φύση.

## **Βιβλιογραφία**

Πληροφορίες για αυτή την ερεύνα μπορείτε να βρείτε στην ιστοσελίδα του Ανδρέα Μέρσιν: http://people.physics.tamu.edu/mershin

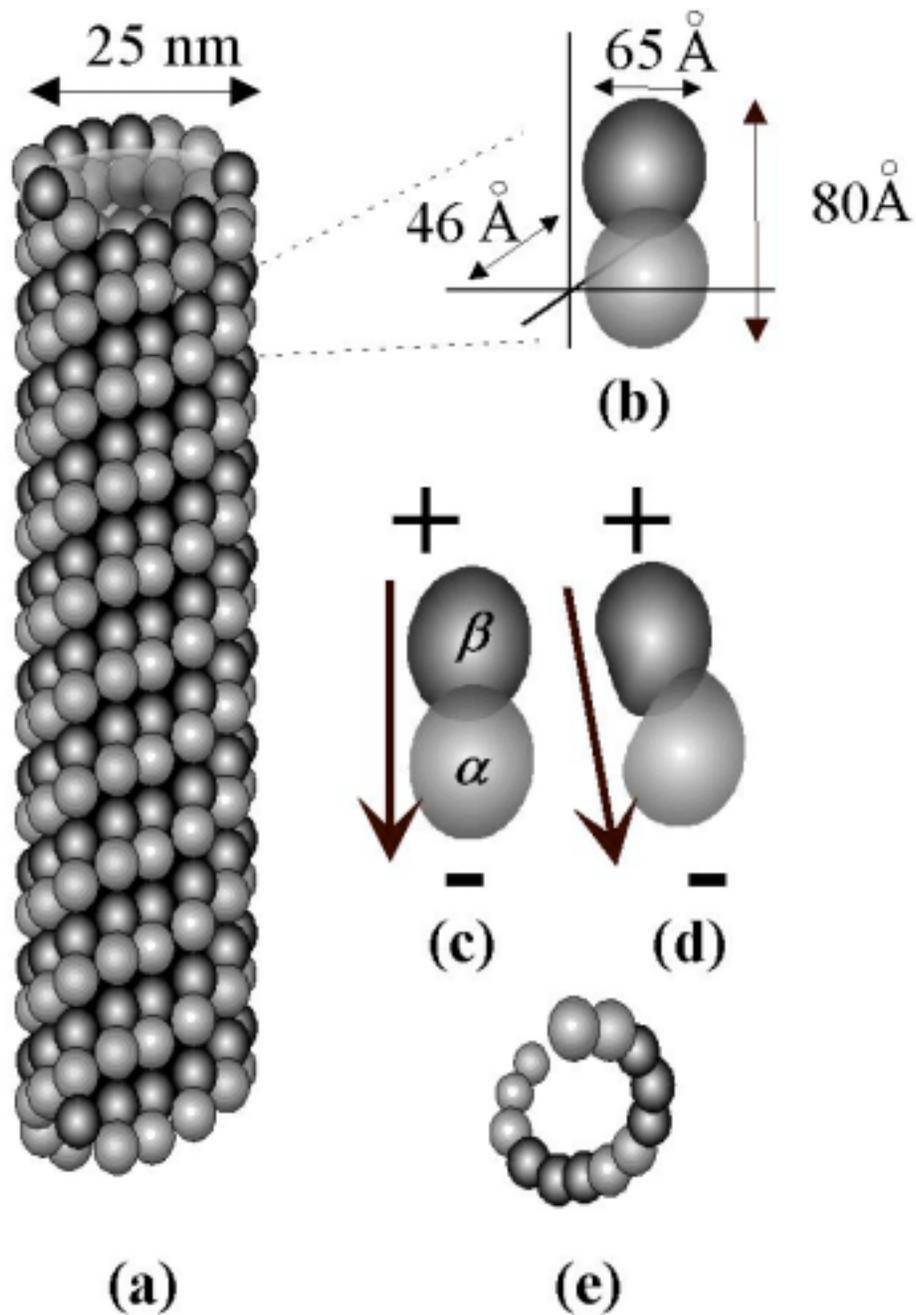

**Εικόνα 1 (a)**: Τυπικό μικροσωληνίδιο αποτελούμενο από 13 νημάτια τουμπουλίνης. **(b)**: Η διαστάσεις του διμερούς τουμπουλίνης. Το μικροσωληνίδιο αποτελείται από **(c)**: GTP-τουμπουλίνη και **(d)**: GDP-τουμπουλίνη. Το βέλος δείχνει την ηλεκτρική διπολική ροπή της κάθε κατάστασης. **(e)**: Εγκάρσια τομή μικροσωληνιδίου που δείχνει το χαρακτηριστικό βήμα. Κάθε νημάτιο είναι μεταθετημένο κατά ένα πέμπτο του ύψους του διμερούς τουμπουλίνης. και έχει βήμα ώστε κάθε 13 τουμπουλίνες συμπληρώνει γωνία 2π ενώ μετατίθεται κάθετα κατά τρεις τουμπουλίνες.

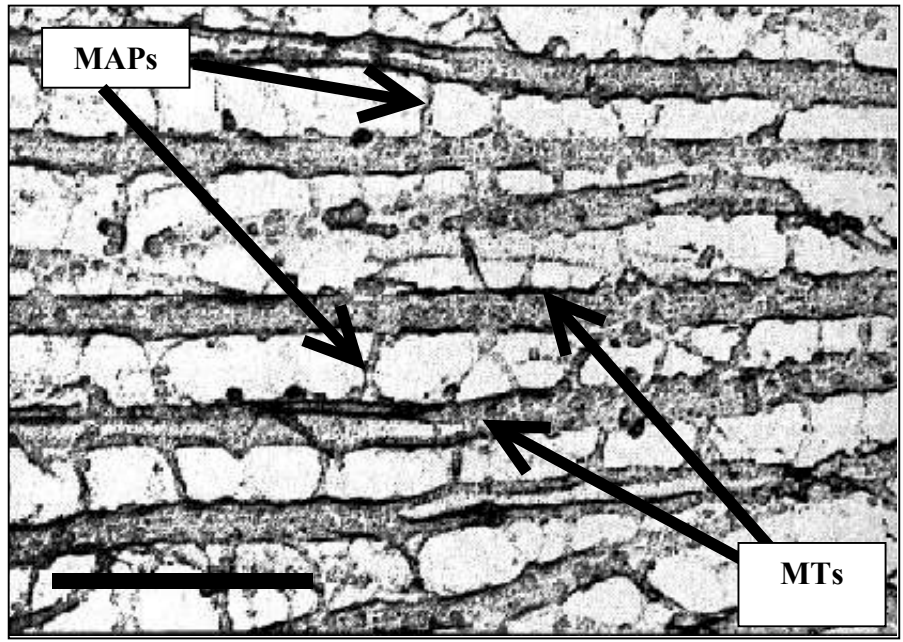

**Εικόνα 2** Φωτογραφία ηλεκτρονικού μικροσκοπίου του κυτταροσκελετού ενός νευρωνικού άξονα. Φαίνονται τα μικροσωληνίδια (MTs) και οι πρωτεΐνες-γέφυρες Microtubule-Associated-Proteins (MAPs). Η μπάρα είναι 100 νανόμετρα. Στο μοντέλο που προτείνουμε η διάταξη των MAPs αντιστοιχεί στο έγγραμμα.